\documentclass[aps,prb,preprintnumbers,twocolumn,showpacs]{revtex4-1}

\usepackage{psfrag,graphicx}
\usepackage{dcolumn}
\usepackage{amsmath,amssymb}
\usepackage{bm}
\usepackage{amsfonts,amssymb,amsmath}        % for math symbols.
\usepackage{multirow}

\usepackage{times}
\usepackage{epstopdf}

\newcommand{\be}{\begin{equation}}
\newcommand{\ee}{\end{equation}}
\newcommand{\bq}{\begin{eqnarray}}
\newcommand{\eq}{\end{eqnarray}}

\newcommand{\rf}[1]{(\ref{#1})}

\graphicspath{{./Figures/}}

\begin{document}

\title{Topological liquid nucleation induced by vortex-vortex interactions in Kitaev's honeycomb model}

\author{Ville Lahtinen$^1$, Andreas W. W. Ludwig$^2$, Jiannis K. Pachos$^3$, and Simon Trebst$^{4,5}$}
\affiliation{$^1$NORDITA, Roslagstullsbacken 23, 106 91 Stockholm, Sweden}
\affiliation{$^2$Department of Physics, University of California, Santa Barbara, CA 93106, USA}
\affiliation{$^3$School of Physics and Astronomy, University of Leeds, LS2 9JT Leeds, UK}
\affiliation{$^4$Microsoft Research, Station Q, University of California, Santa Barbara, CA 93106, USA}
\affiliation{$^5$Institute for Theoretical Physics, University of Cologne, 50937 Cologne, Germany}
\date{\today}

\begin{abstract}

We provide a comprehensive microscopic understanding of the nucleation of topological quantum liquids, a general mechanism where interactions between non-Abelian anyons cause a transition to another topological phase, which we study in the context of Kitaev's honeycomb lattice model. For non-Abelian vortex excitations arranged on superlattices, we observe the nucleation of several distinct Abelian topological phases whose character is found to depend on microscopic parameters such as the superlattice spacing or the spin exchange couplings. By reformulating the interacting vortex superlattice in terms of an effective model of Majorana 
fermion zero modes, we show that the nature of the collective many-anyon state can be fully traced back to the microscopic pairwise vortex interactions. Due to RKKY-type sign oscillations in the interactions, we find that longer-range interactions beyond nearest neighbor can influence the collective state and thus need to be included for a comprehensive picture. The omnipresence of such interactions implies that corresponding results should hold for vortices forming an Abrikosov lattice in a $p$-wave superconductor, quasiholes forming a Wigner crystal in non-Abelian quantum Hall states or topological nanowires arranged in regular arrays.

\end{abstract}


\maketitle

\section{Introduction}
One of the most intriguing aspects of a topological phase is the emergence of anyonic quasiparticles.
If these obey non-Abelian statistics, their presence gives rise to a (macroscopic) ground state degeneracy,
a distinctive feature which has been suggested to be exploited for topological quantum computation.\cite{Nayak08} However,
in any microscopic system this degeneracy will be lifted by the omnipresent
interactions between the anyons. These interactions are often assumed to be extremely weak due to the localized nature of the anyon wavefunction,
but their strength grows exponentially when the anyons are brought into proximity. In fact, they can reach sizable magnitude when the anyon separation becomes of the order of a characteristic length scale, such as the magnetic length in quantum Hall liquids,\cite{Baraban09} 
the coherence length in a $p$-wave superconductor\cite{Cheng09,Sau10}, or the plaquette spacing in Kitaev's honeycomb model.\cite{Lahtinen11}
 
When interacting anyons form regular arrangements 
(e.g. a Wigner crystal in a fractional quantum Hall state \cite{WignerCrystal} or an Abrikosov lattice in a topological superconductor), it has been shown that the collective degeneracy is split, and another topological state
(distinct from the parent state of which the anyons are excitations) is selected as the new ground state.~\cite{Ludwig11} 
This mechanism through which local microscopics can change the global topological properties is referred to as {\sl topological liquid nucleation}. Here we study it in the context of a microscopic model and show that the character of the nucleated topological state can be fully traced back to the signs and the relative magnitudes of the anyon-anyon interactions inherent in the model. This direct connection between {\sl pairwise} interactions and the collective {\sl many-anyon} state is made explicit through an effective Majorana zero mode lattice model for the interacting anyon lattice. Our results provide a comprehensive understanding how nucleation occurs through a hybridization of localized modes and what states can emerge for different anyon lattice densities. Due to the general nature of the employed effective model, our results apply also to microscopically distinct systems supporting localized Majorana mode such as fractional quantum Hall liquids\cite{MooreRead}, $p$-wave superconductors\cite{ReadGreen} and topological nanowires\cite{Sau10,NanoWires}.

\section{Vortex lattices in Kitaev's honeycomb model}

As a prototypical system that allows for good control of all microscopic parameters, we study nucleation in the context of Kitaev's honeycomb lattice model.\cite{Kitaev06} It is an exactly solvable spin model defined by the Hamiltonian
\begin{equation}
   H =  \sum_{{\gamma \rm-links}} { J_{\gamma} \, \sigma_i^{\gamma} \sigma_j^{\gamma}} + K \sum_{(i,j,k)} { \sigma_i^x \sigma_j^y  \sigma_k^z}  \,,
   \label{Eq:KitaevModel}
\end{equation}
where the $\sigma_i^{\gamma}$ denote the standard Pauli matrices describing spin-1/2 moments on the sites of the lattice, $\gamma = x,y,z$ indicates the three different link types, $J_{\gamma}$ are the strengths of the nearest neighbor spin exchange along these links and $K$ is the magnitude of a three spin term that explicitly breaks time reversal symmetry. For a system of $2N$ spins the Hamiltonian \eqref{Eq:KitaevModel} has $N$ local symmetries $[H,\hat{W}_p]=0$, where $\hat{W}_p$ are mutually commuting $Z_2$ valued six-spin operators associated with every plaquette $p$. Their eigenvalues $w_p=-1$ denote a $\pi$-flux vortex at plaquette $p$, while $w_p=1$ denotes an absence of one. One can thus restrict to a particular {\it vortex sector} labeled by the pattern of the eigenvalues $\{ w_p \}$. In each sector the spin model \rf{Eq:KitaevModel} can be mapped to a tight-binding model of free Majorana fermions tunneling on the honeycomb lattice. The corresponding Hamiltonian $H_{\{ w_p\}}$ will always be quadratic in the Majorana operators and thus readily diagonalized.\cite{Lahtinen08}

\begin{figure}[t]	
\includegraphics[width=.9\columnwidth]{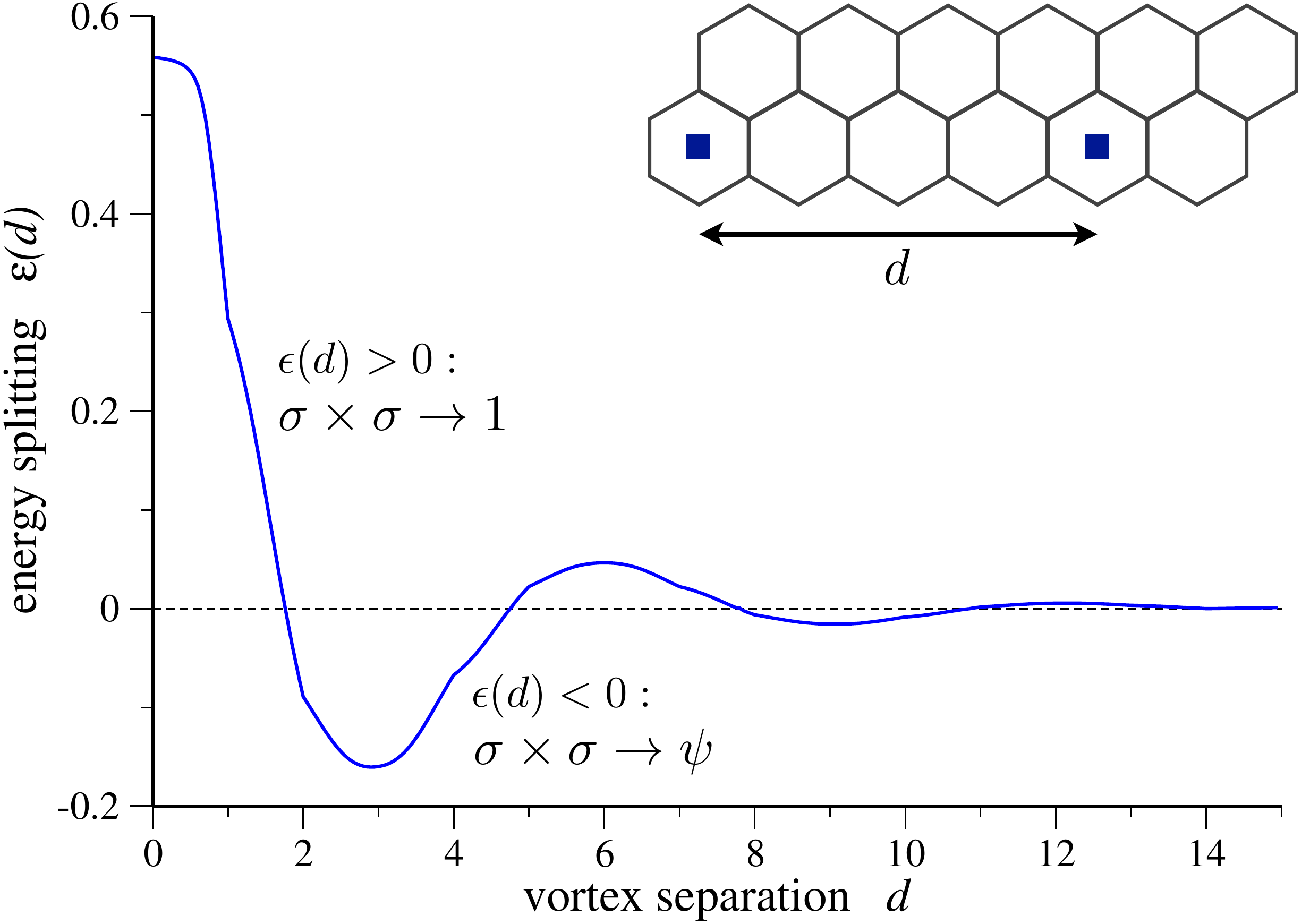} 
\caption{\label{int} Microscopics of the interaction between a pair of vortices in Kitaev's honeycomb lattice model: In agreement with the Ising anyon fusion rule $\sigma \times \sigma = 1 + \psi$, the non-Abelian vortices can combine into two possible collective states with energies $E_1$ and $E_{\psi}$. The splitting $\epsilon(d)=E_\psi-E_1$ between them decreases exponentially with vortex separation and shows characteristic oscillation due to interference effects. When $\epsilon(d)>0$ ($<0$) the microscopics of the system energetically favours the state where the vortices combine into the trivial (fermionic) excitation. The plot is for $J_x=J_y=J_z=1$ and $K=1/20$, which corresponds to a coherence length of $\xi \approx 2.5$. The continuous curve has been obtained by suitably tuning the spin exchange couplings to simulate the continuous transport of the vortices.\cite{Lahtinen11}}
\end{figure}

The ground state over all vortex sectors resides in the vortex-free sector ($w_p=1$ on all plaquettes), which supports phases with both Abelian Toric Code anyons as well as non-Abelian Ising anyons.\cite{Kitaev06} We are interested here in the collective properties of the latter, which appear as magnetic vortex excitations of a phase occurring in the vicinity of isotropic spin exchange $J_x=J_y=J_z$ and for a finite three-spin exchange $K>0$. By studying sectors with only two vortices, it has been shown that the vortices can combine into two distinct collective states whose energies, due to a microscopic interaction, sensitively depend on the separation $d$ between the vortices.\cite{Lahtinen11} The key properties of this pairwise interaction are summarized in Fig.~\ref{int}, which illustrates how the energy splitting $\epsilon(d)$ between the two collective states decays exponentially with increasing separation, {\em i.e.} $\epsilon(d) \sim e^{-d/\xi}$ with $\xi$ being the coherence length of the non-Abelian phase. It further shows RKKY-like sign oscillations at the wavelength of the inverse Fermi momentum. These characteristic features are not unique to the honeycomb model, but they have also been found for Ising anyons emerging in the Moore-Read quantum Hall state,\cite{Baraban09} $p_x + i p_y$ superconductors,\cite{Cheng09} and topological nanowires.\cite{Sau10} The precise values for the amplitude and the frequency of the oscillations, however, depend always on the specific microscopic situation.

The oscillating energy splitting $\epsilon(d)$ is a property of a {\sl two-vortex} problem. Our aim is to use it as an input to understand the {\sl many-vortex} problem and answer which collective ground state is formed as we arrange the interacting vortices in superlattices. In the honeycomb model this corresponds to studying vortex sectors where the eigenvalues $w_p=-1$ form a periodic pattern. For simplicity we restrict to considering uniform and isotropic superlattices, which, as illustrated in Fig.~\ref{vlattice}, we can parametrize with the superlattice spacing $D=1,2,3,\ldots$ (in units of the plaquette spacing). All these sectors are translationally invariant with respect to a suitably chosen magnetic unit cell, with the corresponding Bloch Hamiltonians $H_{D=\{w_p\}}$ being readily (numerically) diagonalizable $4D^2 \times 4D^2$ matrices.\cite{Lahtinen08}

\section{Nucleated phases and vortex band structure}

Varying the spacing $D$ is of interest for two physical reasons. First, it enables us to tune the microsccopic pairwise interactions, as captured by the energy splittings $\epsilon(D)$, which we find resulting in different collective topological states. We characterize them by the Chern number\cite{Kitaev06} that can be numerically evaluated for the ground state of each sector.\cite{Fukui05} Second, going systematically through the sequence of possible superlattice spacings enables us to simulate the expansion of a Wigner crystal of quasiholes when shifting the magnetic field on a quantum Hall plateau, or similarly the expansion of an Abrikosov lattice of vortices in a $p$-wave superconductor when varying the applied magnetic field. 

\begin{figure}[t]
\includegraphics[width=\columnwidth]{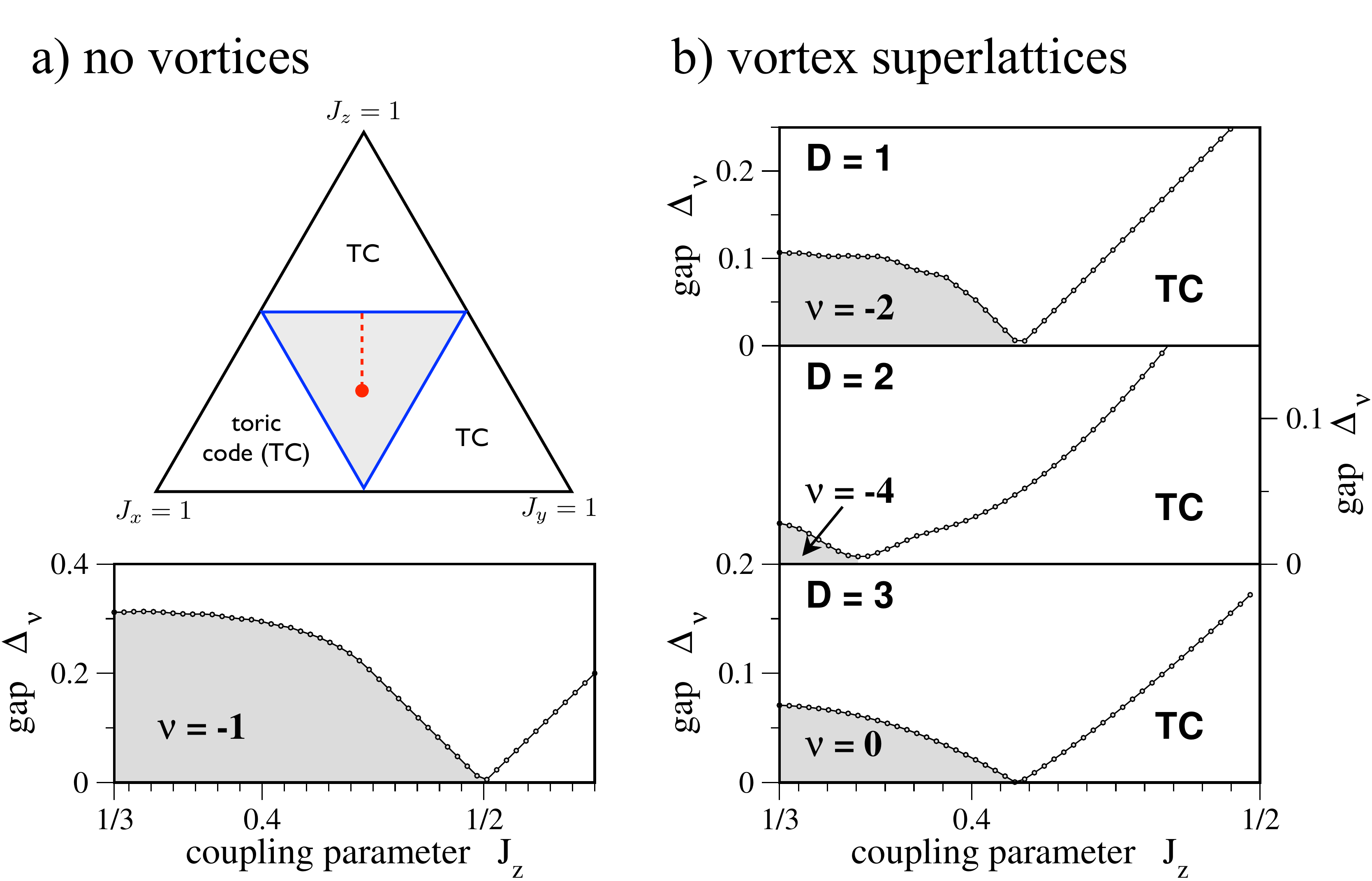}
\caption{\label{PD} 
   	Phase diagram of Kitaev's honeycomb model (a) in the absence of a vortex superlattice and (b) in the presence of $D=1,2,3$ superlattices. We have
	parametrized the phase diagram by $\sum_\gamma J_\gamma = 1$ and set $K=0.03$. The plots show the lowest energy gap $\Delta_v$ and Chern numbers $\nu$ along a cut ($1/3 \leq J_z \leq 1/2$ when $J_x=J_y$) indicated by the dashed red line in (a). In the absence of vortices the continuous phase transition between the Abelian toric code phase ($\nu=0$, denoted by TC) and the non-Abelian Ising phase ($\nu=-1$) occurs at $J_z=1/2$. In the presence of a vortex superlattice we again find a continuous phase transition, but which occurs now always at some $J_z^c<1/2$ separating the TC phase ($J_z>J_z^c$) from the nucleated Abelian phase ($J_z<J_z^c$). The exact value of $J_z^c(D)$ can also be traced back to the vortex-vortex interactions, with there being no simple direct relation between the spacing $D$ and the critical value.\cite{Lahtinen_inprep}}
\end{figure}

Our findings are given in Fig.~\ref{PD}, which shows two general ways the presence of a vortex superlattice modifies the phase diagram. First, the non-Abelian phase (characterized by Chern number $\nu=-1$) occurring in the vicinity of the isotropic spin exchange $J_x=J_y=J_z$ is always replaced by a distinct {\sl Abelian} topological phase (characterized by an even Chern number). In particular, we find phases characterized by $\nu = -2, -4, 0$ for integer superlattice spacings $D=1,2,3$, respectively. Second, we find that the Abelian phases existing in the dimerized limits (e.g. $J_z > J_x+J_y$), that are characterized by $\nu=0$ and support non-chiral Abelian anyons (so-called semions) \cite{Kitaev06,Vidal08}, are always enlarged in the presence of a vortex superlattice. We focus here to understand the first phenomenon and show that all the Abelian phases emerging in the vicinity of $J_x=J_y=J_z$ are {\it nucleated phases}, i.e. they arise due to the vortex-vortex interactions. 
%The anyonic excitations in these phases are vortex vacancies, whose properties follow from the Chern number characterizing the ground state.\cite{Kitaev06} 

\begin{figure}
\includegraphics[width=\columnwidth]{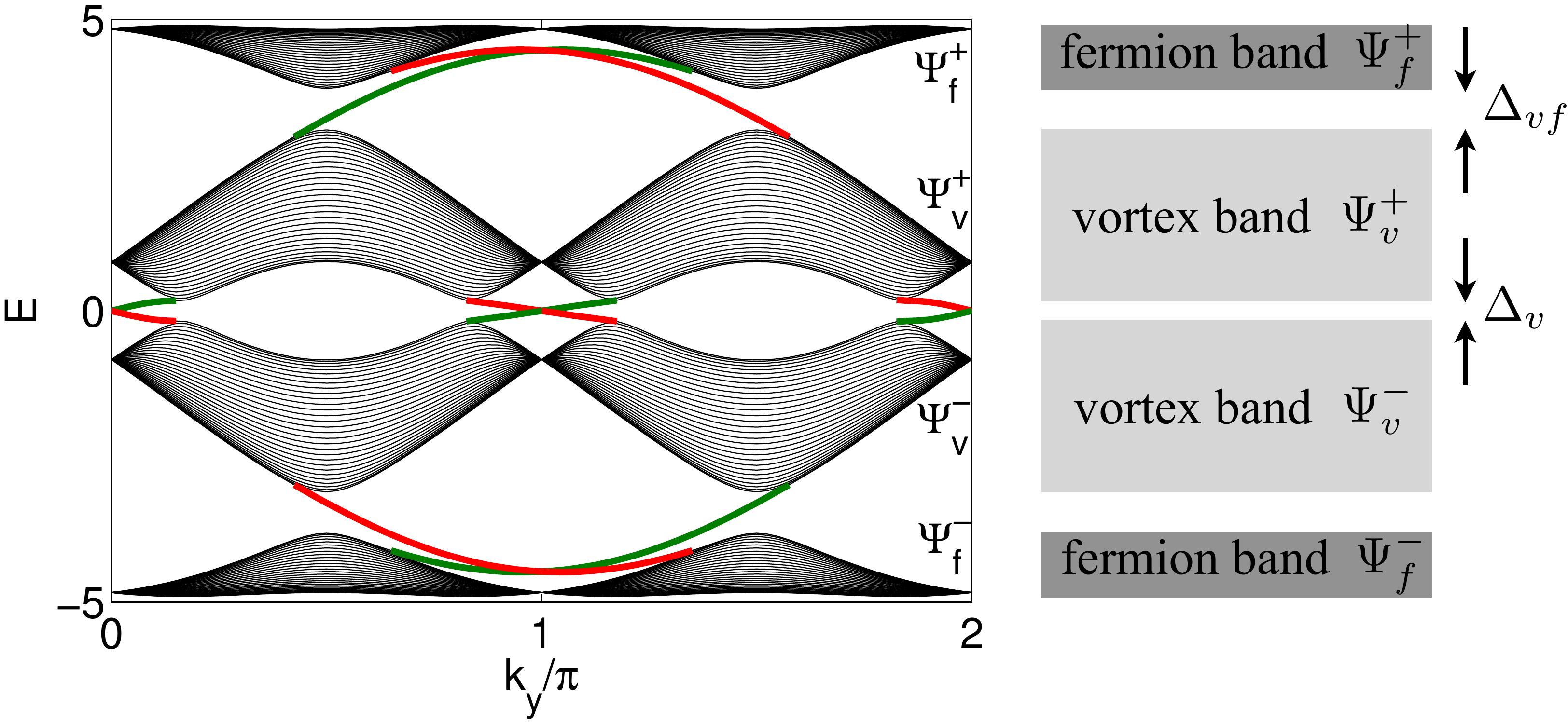} 
\caption{\label{edges} The characteristic band structure of the nucleated topological phases with
					 $\Psi_v^\pm$ ($\Psi^\pm_f$) denoting the low-energy (high-energy) vortex (fermion) bands.
					 The actual plot is for the $D=1$ vortex superlattice on a cylinder (open boundary conditions in $x$-direction). Consistent with the nucleated Chern number $\nu=-2$ phase, the spectral flow shows two edge modes per edge (red and green lines denoting different edges) emanating from the four gapped Dirac cones.
}
\end{figure}

A further investigation of the nucleated phases reveals characteristic band structures. The spectrum of the $\nu=-2$ phase arising in the presence of a full-vortex ($D=1$) superlattice, shown in Fig.~\ref{edges}, exhibits four gapped Dirac cones.\cite{Lahtinen10} The spectrum of the $\nu=-4$ phase arising in the presence of a $D=2$ superlattice also exhibits four gapped Fermi points, but with a significantly broader (quadratic) dispersion. The energy spectrum of the nucleated phase for the $D=3$ superlattice, however, does not show indications of Fermi points in agreement with the phase being characterized by $\nu=0$. Unlike the other nucleated phases, this phase remains gapped even as the three-spin coupling $K$ is tuned to zero. This indicates that it is adiabatically connected to the previously observed gapped phase,~\cite{Kamfor11} that emerges when a vortex superlattice is imposed on the {\sl gapless} time-reversal symmetric spin liquid (for which $K=0$). Despite this behavior suggesting a different origin (the non-Abelian anyons underlying the nucleated phases emerge only for $K \neq 0$), we will show below that also this phase can be traced back to the vortex-vortex interactions.

The crucial common feature of all these observed band structures is that they consist of a low-energy \emph{vortex band} $\Psi_v^\pm$ and a set of high-energy \emph{fermion bands} $\Psi_f^\pm$ (see Fig.~\ref{edges} for an illustration). In the presence of $2N$ vortices the first contains $N$ states that have support only on the sites near the vortices, while the latter contains the rest of the states that in general have support on all sites of the honeycomb lattice. For any finite $K$ the vortex and fermion bands are separated in energy by a band gap $\Delta_{vf}$. Hence the Chern number characterizing the ground state can be written as
\be \label{nu}
	\nu = \nu_v + \nu_f,
\ee
where $\nu_v$ and $\nu_f$ are the Chern numbers for the occupied negative energy bands $\Psi_v^-$ and $\Psi_f^-$ bands, respectively. We observe that the first depends on the underlying vortex configuration, while the latter contributes always $\nu_f=-1$ for $K>0$. The nucleated phases can thus be viewed as comprising of two decoupled theories: a remnant of the non-Abelian phase living on the honeycomb lattice and an emergent theory living effectively ``on top of it'' on the vortex lattice. The problem of understanding the nature of the collective many-vortex state is therefore reduced to the problem of understanding how $\nu_v$ depends on $D$.

\section{Effective Majorana model for the interacting vortex superlattice}

\begin{figure}
\begin{tabular}{cc}
\includegraphics[width=4.2cm]{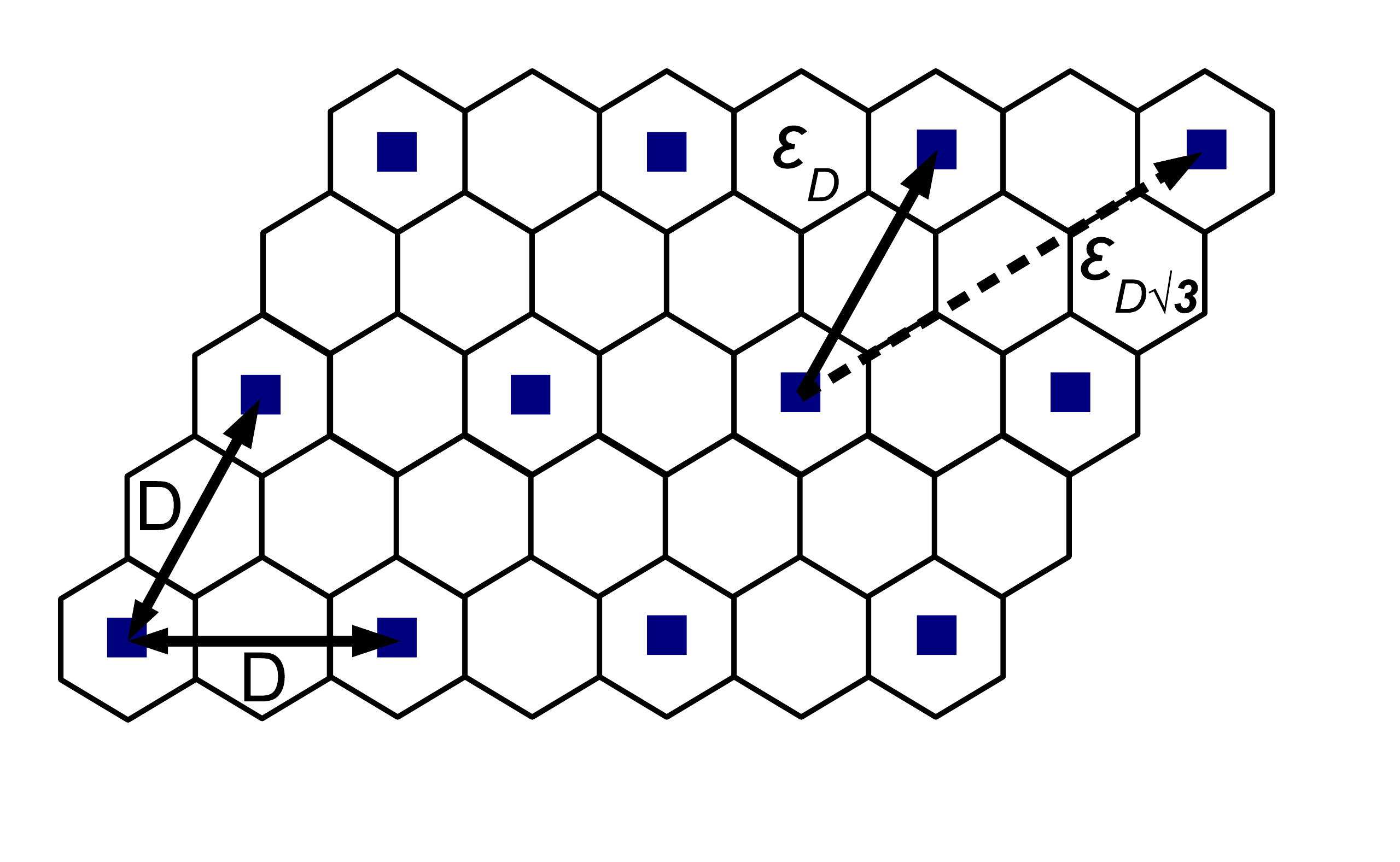} & \includegraphics[width=4.2cm]{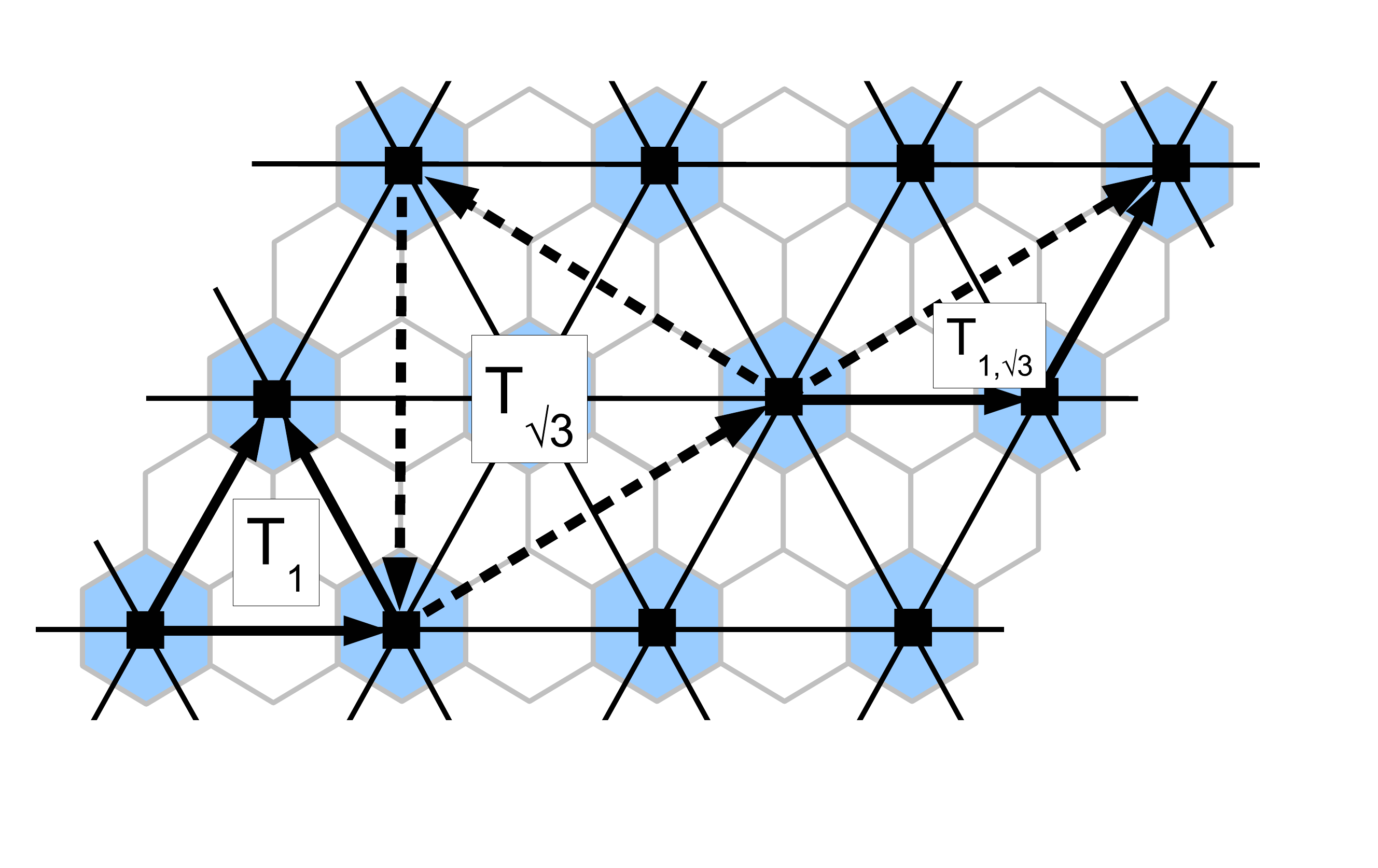} \\
(a) & (b)
\end{tabular}
\caption{\label{vlattice} (a) A vortex (blue squares) superlattice of spacing $D$ (here $D=2$) and
 (b) the corresponding effective model in terms of Majorana zero modes tunneling on the triangular superlattice. The effective hopping amplitudes $t_1$ (solid lines) and $t_{\sqrt{3}}$ (dashed lines) are identified with the energy splittings $\epsilon(D)$ and $\epsilon(D\sqrt{3})$ at the corresponding vortex separations, respectively.
       The fluxes in the effective model are chosen such that plaquettes indicated as $T_1$, $T_{\sqrt{3}}$, and $T_{1,\sqrt{3}}$
        are assigned fluxes $\Phi_{T_1} =\frac{\pi}{2}$, $\Phi_{T_{\sqrt{3}}} = -\frac{\pi}{2}$, 
        and $\Phi_{T_{1,\sqrt{3}}} = \frac{\pi}{2}$, respectively, that are consistent with the enclosed area.
     }
\end{figure}

We now turn to address the microscopic connection between the presence of a given vortex superlattice and the nature of the nucleated phase.
To this end we consider an effective tight-binding model of Majorana fermions that connects the energy splitting $\epsilon(d)$ due to vortex-vortex interactions to the Chern number $\nu_v$ of the vortex band characterizing the nucleated topological phase. The motivation for such a model is as follows: The vortices bind localized Majorana zero modes and the oscillating energy splitting shown in Fig.~\ref{int} can be viewed as arising from them tunneling between the two vortex cores.\cite{Lahtinen11} Likewise, we assume that the vortex band $\Psi_v^-$ can be understood as arising through a collective hybridization, which occurs as the Majorana zero modes start tunnelling on the triangular vortex superlattice. To capture also the case when longer range interactions become relevant our model incorporates both nearest (n.n.) and next-nearest (n.n.n.) neighbor hopping of amplitudes $t_1$ and $t_{\sqrt{3}}$, respectively (see Fig.~\ref{vlattice}). The effective Majorana Hamiltonian is then given by
\begin{equation}
   H_M = it_1 \sum^{\rm n.n.}_{\langle ij \rangle} s_{ij} \, \gamma_i \gamma_j + it_{\sqrt{3}} \sum^{\rm n.n.n.}_{\langle\langle ij \rangle\rangle} s_{ij} \, \gamma_i \gamma_j \,,
   \label{Eq:MajoranaModel}
\end{equation}
where the $\gamma_i$ denote Majorana zero modes at vortex location $i$ (the center of honeycomb plaquette) obeying $\{ \gamma_i, \gamma_j \} = 2 \delta_{ij}$.
The model allows for $Z_2$ gauge choices $s_{ij}=\pm1$, which give rise to flux $\Phi_{ijk} = -i \ln ( i s_{ij} s_{jk} s_{ki} ) = \pm \pi/2$ through each triangular plaquette with corners $i,j,k$. As illustrated in Fig.~\ref{vlattice}, there are three distinct types of plaquettes: those spanned solely by either $t_1$- or $t_{\sqrt{3}}$-links and those consisting of both. We denote the former as $T_1$ and $T_{\sqrt{3}}$, respectively, and the latter as $T_{1,\sqrt{3}}$. When $t_{\sqrt{3}}=0$, it has been shown that for a uniform triangular lattice of $\pi$-flux vortices one should impose a $+\pi/2$- or $-\pi/2$-flux on all triangles $T_1$.\cite{Grosfeld} In the following we will fix this flux to be $\Phi_{T_1}=+\pi/2$. 
On the triangles $T_{\sqrt{3}}$ and $T_{1,\sqrt{3}}$, which involve the $t_{\sqrt{3}}$ hopping, we fix the fluxes to be $T_{\sqrt{3}} = 3\pi/2 = -\pi/2$ (mod $2\pi$) and $\Phi_{T_{1,\sqrt{3}}} = +\pi/2$ such that the flux through each triangle is proportional to its enclosed area. A periodic pattern of gauge choices $s_{ij}$ satisfying these flux assignments requires a magnetic unit cell of at least 36 sites (see Appendix A). 

\begin{figure}
  \includegraphics[width=\columnwidth]{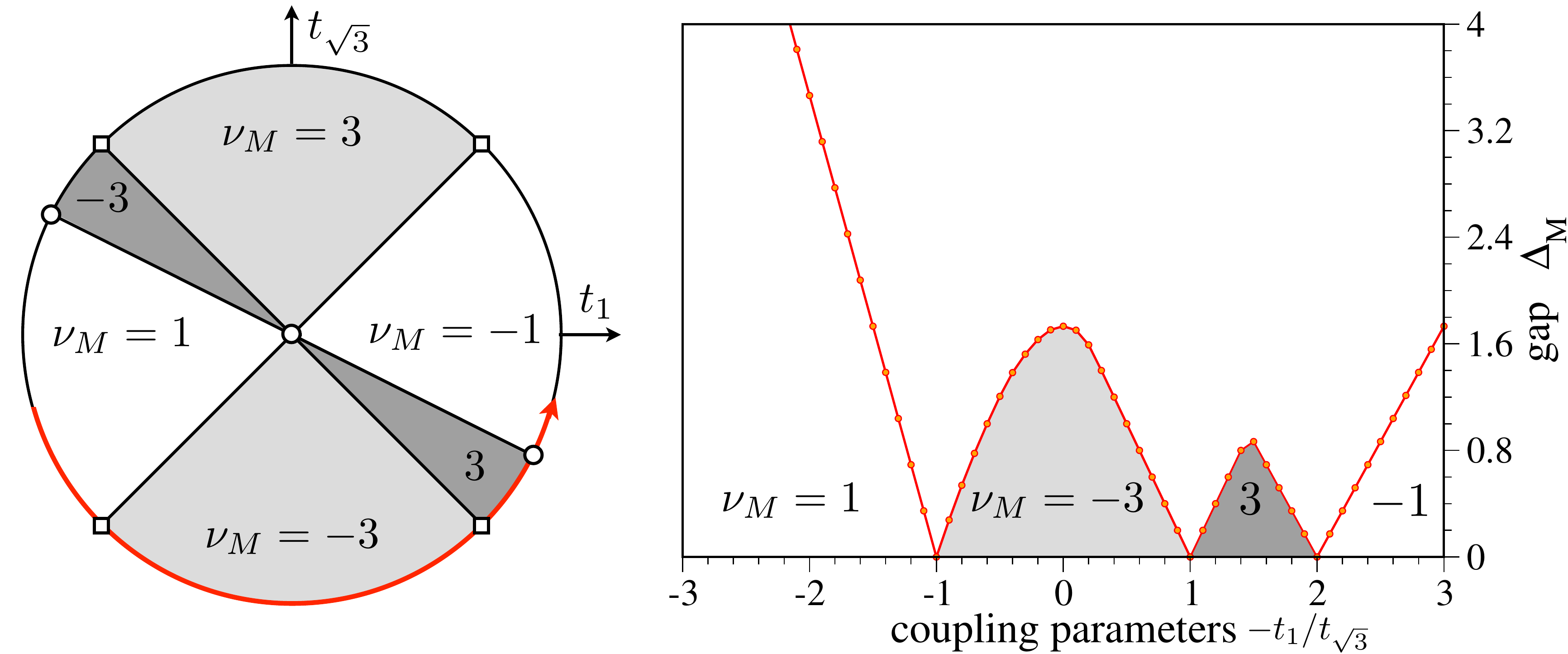}
\caption{\label{eff_pd} Left: The phase diagram of the effective Majorana fermion zero mode model (here $t_1 = \cos{\theta}$  and $t_{\sqrt{3}} = \sin{\theta}$), as characterized by the Chern numbers $\nu_M$, for the flux assignment $(\Phi_{T_1},\Phi_{T_{\sqrt{3}}},\Phi_{1,T_{\sqrt{3}}})=(\frac{\pi}{2},-\frac{\pi}{2},\frac{\pi}{2})$. The squares and the circles denote phase transitions at $|t_1|=|t_{\sqrt{3}}|$ and $t_1=-2t_{\sqrt{3}}$, respectively. Right: The lowest gap in the energy spectrum, $\Delta_M$, along the path shown in (a) for $t_{\sqrt{3}}=-1$. }
\end{figure}

We (numerically) diagonalize this effective model for varying relative amplitudes of the two hopping terms $t_1$ and $t_{\sqrt{3}}$. Fig.~\ref{eff_pd} illustrates the resulting phase diagram that shows various gapped phases characterized by Chern numbers $\nu_M= \pm 1,\pm 3$.
The occurrence of some of these phases can be readily understood in the following way: If the nearest-neighbor hopping dominates
$(t_1 \gg t_{\sqrt{3}})$, we recover the analytically tractable triangular lattice problem studied before \cite{Grosfeld}, which has been shown
to give rise to a gapped phase with $\nu_M = \pm 1$. Similarly, in the opposite limit ($t_{\sqrt{3}} \gg t_1$) the system simply 
decomposes into three uncoupled copies of the triangular lattice problem, thus giving a Chern number of $\nu_M = \pm 3$. There are also intermediate
phases with Chern number $\nu_M = \pm 3$ that arise for $-2 < t_1/t_{\sqrt{3}} < -1$ and arise due to a competition between the two limiting cases above. The general symmetry of the phase diagram where $\nu_M \to -\nu_M$ as $t_l \to -t_l$ is characteristic for the triangular lattice, where the elementary plaquettes are odd cycles. Inverting the signs of all tunneling amplitudes is equivalent to inverting flux $\pm \pi/2 \to \mp \pi/2$ on all plaquettes, which leads to a time reversed phase with a Chern number of opposite sign.   

Equipped with these quantitative results for the effective Majorana model, we return to establishing a direct connection between the microscopic pairwise vortex interactions and the collective ground state in the presence of a spacing $D$ superlattice of interacting vortices. The interactions enter \eqref{Eq:MajoranaModel} by identifying the hopping amplitudes $t_1$ and $t_{\sqrt{3}}$ with the corresponding energy splittings $\epsilon(D)$ and $\epsilon(D\sqrt{3})$, respectively, as illustrated in Fig.~\ref{vlattice}. More precisely, we will use the ansatz
\be \label{tl}
	t_l = (-1)^{P_{lD}} |\epsilon(lD)| \,,
\ee
where $P_d$ is the fermionic parity ($P_d=0$ for even, $P_d=1$ for odd) of the respective two vortex sector with vortex separation $d$. The fermionic parity determining the sign of the physical energy splitting originates from the mapping of the spin Hamiltonian \rf{Eq:KitaevModel} to Majorana Hamiltonian on the honeycomb lattice\cite{Kells09} and it is thus specific only to the honeycomb model. In the other microscopic models with interacting anyons both the magnitude and the sign can be obtained directly from the oscillating energy splitting.

\section{Results}

To show that our effective model correctly predicts the Chern numbers of the nucleated phases, we will restrict to isotropic $J_x=J_y=J_z$ couplings where the Hamiltonian \rf{Eq:KitaevModel} has rotational $C_3$ symmetry. Anisotropic couplings lead to anisotropic interactions, which require a more complicated effective model with correspondingly anisotropic hopping amplitudes.\cite{Lahtinen_inprep} Thus, for simplicity we restrict ourselves to tuning only the magnitude $K$ of the three spin term. It respects the rotational symmetry and the isotropic effective model \rf{Eq:MajoranaModel} is valid for all values of it. The idea is then as follows: By tuning the microscopic coupling $K$ we can tune the honeycomb model in the presence of a vortex superlattice through various topological phases. Independently this will tune also the pairwise energy splittings $\epsilon(d)$ to different microscopic values, which in turn through \rf{tl} will tune the state of the effective Majorana model. If the interactions are indeed responsible for the emergent Abelian phases, we expect agreement between the observed ($\nu$) and predicted ($\hat{\nu}_M=\nu_M-1$) Chern numbers. Furthermore, we expect the interaction induced nucleated gap ($\Delta_M$) to approximate the observed energy gap ($\Delta_v$).

\begin{figure}[t]
\includegraphics[width=\columnwidth]{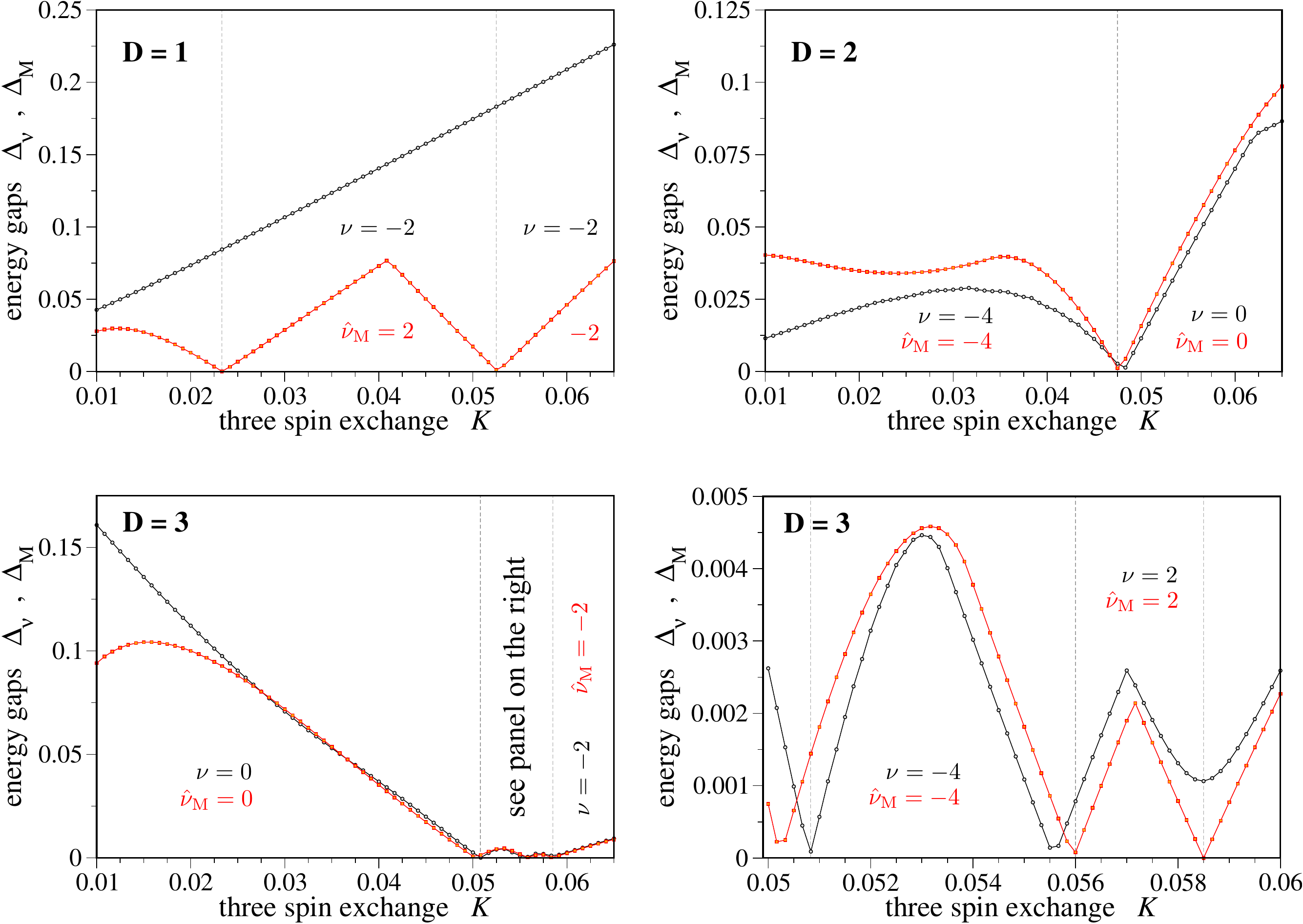} 
\caption{\label{PDK} The energy gaps $\Delta_v(K)$ of the nucleated phases in the honeycomb model (black circles) and the gaps $\Delta_M[\epsilon(D,K),\epsilon(D\sqrt{3},K)]$ predicted by the effective model (red squares) for $D=1,2$ and $3$ vortex superlattices. $\nu$ is the Chern number observed in the honeycomb model and $\hat{\nu}_M=\nu_M-1$ is the one predicted by the Majorana model from the interactions. Note the quantitative similarity of the sequence of phases for $D=3$ superlattice to that of Fig.~\ref{eff_pd} (the finite gap around $K=0.0585$ is a finite-size effect). All the data is for $J_x=J_y=J_z=1/3$ and the values of the pairwise energy splittings $\epsilon(d)$ have been obtained by restricting to the relevant two vortex sectors in a finite system of $40 \times 40$ plaquettes (3200 sites) on a torus.}
\end{figure}

Our results are summarized in Fig.~\ref{PDK}, which indeed shows {\it quantitative} agreement between the observed and predicted energy gaps and Chern numbers. This is most clearly illustrated for the $D=3$ superlattice, where the observed sequence of phases with Chern numbers $\nu=0,-4,2,-2$ matches exactly that predicted by the effective model. In general, the approximation provided by the effective model gets more accurate for larger $K$ and sparser vortex superlattices (for $D=1$ the Chern numbers agree only for large $K$, while for $D>2$ they always agree). The reason is that for small $K$ or for tight superlattices the coherence length $\xi$ of the underlying non-Abelian phase can increase (as the gap $\Delta \sim K^{-1}$ decreases) beyond the superlattice spacing \cite{Lahtinen11}. This presumably renders the notion of individual vortices poorly defined and thus makes our microscopic approach inapplicable (while nucleation still occurs, the tunneling amplitudes are no longer captured by \rf{tl} and/or many-body effects become relevant). Indeed, even if the nucleated gap $\Delta_v$ decays exponentially with increasing $D$, in agreement with it being induced by the interactions, further studies for spacings up to $D=6$ shows excellent agreement over a wide range in $K$. 

Finally, we can now understand that the phases with Chern number $\nu=-4$ and $2$ emerge due to the longer range interactions. Even if they are exponentially suppressed, the oscillations can cause nearest neighbor interactions to vanish and thus make the next nearest interactions to dominate the physics. We could have included in the model interactions of even longer range, but as \rf{Eq:MajoranaModel} can account for all the observed Chern numbers, we regard it as providing a complete description. The exponential decay of the interactions means that if stable phases due to even longer range interactions existed, they should have appeared for the considered dense superlattices.

\section{Conclusions} 

We have studied in microscopic detail the nucleation of topological liquids -- a general mechanism where interactions between non-Abelian anyons cause a transition to another topological phase\cite{Ludwig11} -- in the context of Kitaev's honeycomb model. Generally, we find that the presence of a superlattice of interacting Ising vortices always destroys the non-Abelian nature of the underlying topological phase and nucleates an Abelian topological phase. Employing an effective model of Majorana zero modes tunneling on the vortex lattive, we show that the character of the collective {\sl many-vortex} state and the energy gap protecting it can be fully traced back to the sign, amplitude and ratio of the {\sl pairwise} vortex-vortex interactions of different range. This provides an explicit demonstration of how local microscopics can change the global topological nature of the system.

We found Abelian phases with Chern numbers $\nu = +2$ or $\nu = -4$ that arise when oscillations in the interactions, even if they decay exponentially, cause next-nearest neighbor interactions to become of comparable (but of different sign) or larger magnitude than the nearest-neighbor interactions. Due to the omnipresence of such oscillating anyon-anyon interactions,\cite{Baraban09,Cheng09,Sau10} our results imply that a similar nucleation mechanism should occur also in other microscopically distinct topological systems. These include Moore-Read quantum Hall states, where quasihole excitations can form a Wigner crystal when the magnetic field is detuned away from the middle of the plateau,\cite{WignerCrystal} $p$-wave superconductors, where vortices can form an Abrikosov lattice, and topological nanowires arranged in regular arrays.\cite{NanoWireArray} While we focused here on nucleation in the presence of an ideal vortex superlattice, further studies\cite{Lahtinen_inprep} will address the microscopics of a recently observed thermal metal state in the presence of disorder.\cite{Laumann11}

\section*{Acknowledgements} 
This work has been supported, in part, by  the Finnish Academy of Science (V.L),
 the NSF (A.W.W. L.) through DMR-0706140, and the Royal Society (J.P.),
V.L. would like to thank G. Kells for insightful discussions. 
S.T. acknowledges hospitality of the Aspen Center for Physics.

\appendix

\section{Unit cell for the Majorana model with longer range tunneling}

\begin{figure}[t]
\includegraphics[width=\columnwidth]{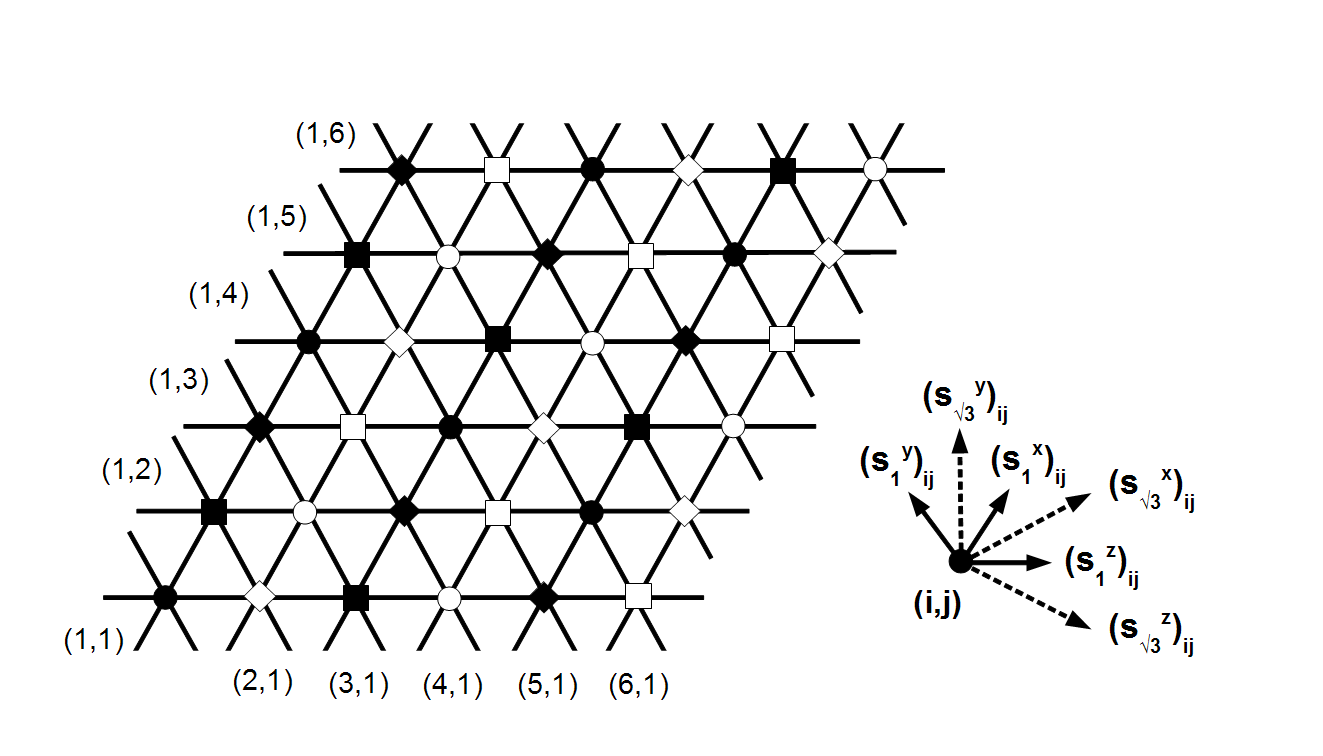} 
\caption{\label{unitcell} {\it Left:} The unit cell for the effective Majorana model with both nearest ($t_1$) and next nearest ($t_{\sqrt{3}}$) neighbor tunneling. The first couple all sites of the lattice, while the latter couple only sites belonging to one of three distinct sublattices denoted by circles, squares and diamonds. When $t_{\sqrt{3}}=0$ the unit cell consists of a black and a white site irrespective of their sublattice label, while for $t_1=0$ the unit cell consists of six sites -- a black and a white site from each sublattice. {\it Right:} For every site $(i,j)$ in the unit cell one has six independent gauge choices --  three on the $t_1$ lattice and three on the respective $t_{\sqrt{3}}$ sublattice. When calculating the flux per plaquette, one has to take into account the overall $i$ factor in the tunneling. We assume a convention that the phase of the hopping between any two sites is given by $i$ ($-i$) when it is along (against) the shown orientations.}
\end{figure}

When $t_{\sqrt{3}}=0$ the unit cell with $\Phi_{T_1}=\pi/2$-flux on all plaquettes contains two sites. In the other limiting case when $t_1=0$, the unit cell with $\Phi_{T_{\sqrt{3}}}=-\pi/2$ consists of six sites, two each from each of three disjoint sublattices on which the next nearest hopping act. When both are present, one needs also to fix the flux on all the intermediate plaquettes, that consist of two $t_1$-links and a single $t_{\sqrt{3}}$-link, to be $\Phi_{T_{1,\sqrt{3}}}=\pi/2$. It turns out that a gauge consistent with all these flux assignments is translationally invariant only with respect to a magnetic unit cell of 36 sites, as illustrated in Fig.\ref{unitcell}.

For each site $(i,j)$ in the unit cell one has six independent gauge choices -- $(s^x_1)_{ij}$,$(s_1^y)_{ij}$ and $(s_1^z)_{ij}$ on the $t_1$-lattice and $(s^x_{\sqrt{3}})_{ij}$,$(s_{\sqrt{3}}^y)_{ij}$ and $(s_{\sqrt{3}}^z)_{ij}$ on the $t_{\sqrt{3}}$-lattice. Taking into account the orientation of the links when evaluating the flux (see Fig.\ref{unitcell}), one possible gauge choice giving rise to the desired $(\Phi_{T_1},\Phi_{T_{\sqrt{3}}},\Phi_{1,T_{\sqrt{3}}})=(\pi/2,-\pi/2,\pi/2)$ flux pattern is given by
\bq
 s_1^x & = & \left( \begin{array}{cccccc}
 -1 & 1 & -1  &  1  &  1 & 1 \\
-1  &  1 &  -1  &  1  & -1 & -1 \\
 1  &  1 &  -1  &  1  & -1 & 1 \\
-1  & -1 &  -1  &  1  & -1 & 1 \\
-1  &  1 &   1  &  1  & -1 & 1 \\
-1  &  1 &  -1  & -1  & -1 & 1
\end{array} \right) \nonumber \\
s_1^y & = & \left( \begin{array}{ccccccc} 
		-1  &  1 &  -1  &  -1  &  -1 &  1 \\
    -1  &  1 &  -1  &  1  & 1 & 1 \\
     -1  &  1 &  -1  &  1  & -1 &  -1 \\
    1  & 1 &  -1  &  1  & -1 &  1 \\
    -1  &  -1 &   -1  &  1  & -1 &   1 \\
    -1  &  1 &  1  & 1  & -1 &  1 
\end{array} \right) \nonumber \\
s_1^z & = & \left( \begin{array}{ccccccc} 
		1  &  1 &  1  &  1  & -1 & -1 \\
    -1  &  1 &  1  &  1  & 1 & -1 \\
     -1  &  -1 &  1  &  1  & 1 &  1 \\
    1  & -1 &  -1  &  1  & 1 &  1 \\
    1  &  1 &   -1  &  -1  & 1 &   1 \\
    1  &  1 &  1  & -1  & -1 &  1 
\end{array} \right) \nonumber
\eq
and
\bq
 s_{\sqrt{3}}^x & = & \left( \begin{array}{cccccc}
 1 & 1 & -1  &  1  &  1 & -1 \\
1  &  -1 &  -1  &  1  & -1 & -1 \\
 1  &  -1 &  1  &  1  & -1 & 1 \\
-1  & -1 &  1  &  -1  & -1 & 1 \\
-1  &  1 &   1  &  -1  & 1 & 1 \\
-1  &  1 &  -1  & -1  & 1 & -1
\end{array} \right) \nonumber \\
s_{\sqrt{3}}^y & = & \left( \begin{array}{ccccccc} 
		-1  &  -1 &  -1  &  -1  &  -1 &  -1 \\
    -1  &  -1 &  -1  &  -1  & -1 & -1 \\
     -1  &  -1 &  -1  &  -1  & -1 &  -1 \\
    -1  & -1 &  -1  &  -1  & -1 &  -1 \\
    -1  &  -1 &   -1  &  -1  & -1 &   -1 \\
    -1  &  -1 &  -1  & -1  & -1 &  -1 
\end{array} \right) \nonumber \\
s_{\sqrt{3}}^z & = & \left( \begin{array}{ccccccc} 
		-1  &  -1 &  1  &  -1  & -1 & 1 \\
    -1  &  1 &  1  &  -1  & 1 & 1	 \\
     -1  &  1 &  -1  &  -1  & 1 &  -1 \\
    1  & 1 &  -1  &  1  & 1 &  -1 \\
    1  &  -1 &   -1  &  1  & -1 &   -1 \\
    1  &  -1 &  1  & 1  & -1 &  1 
\end{array} \right) \nonumber
\eq

\end{document}